\documentclass[a4paper,fleqn]{article}
\usepackage{amsmath}
\usepackage{times}
\usepackage{graphicx,times}
\title{\flushleft\textbf{A conjecture on Kepler's third law of n-body periodic orbits}}

\author{\begin{small}Chang-Yin Zhao$^{1,2}$\footnote{Corresponding author, E-mail:
cyzhao@pmo.ac.cn}\quad and Ming-Jiang Zhang$^{1,2}$\end{small}\\
\begin{small} $^{1}$Purple Mountain Observatory, Chinese Academy of Sciences, Nanjing 210034, China\end{small}\\
\begin{small} $^{2}$Key Laboratory of Space Object and Debris Observation, PMO, CAS, Nanjing 210034, China \end{small}}
\date{\today}

\begin{document}
\maketitle
\noindent\textbf{\emph{\textbf{Abstract}}} \emph{Three-body and n-body problems in celestial mechanics are age-old and challenging puzzles. In recent years, several breakthroughs are made in finding periodic orbits for three-body problem. And Bohua Sun proposed a conjecture on Kepler's third law of three-body and n-body problems by using the dimensional analysis method and the mass product symmetry of Newtonian gravitational field. In this paper, the background as well as the research progress on the Kepler's third law, three-body and n-body problems is introduced briefly, and then Bohua Sun's conjecture on Kepler's third law of three-body and n-body problems is reviewed from the perspective of both theory and application.}\\
\\
\textbf{PACS numbers:} 95.10.Ce, 45.50.Pk\\
\textbf{Key words:} celestial mechanics, three-body problem, n-body problem, Kepler's third law

\section*{}
\hskip\parindent In 1619, ten years after the publishing of his first two laws, Kepler proposed his third law, which captured the relationship between the period $T$ and the semi-major axis $a$ of the motion of celestial bodies \cite{1}: $T^2/a^3=K$, where $K$ is a constant. Kepler found these by analysing the astronomical observations of Tycho Brahe, rather than by way of mathematical deduction. Newton, who discovered the universal law of gravitation, found that the planets' orbits conform approximately to Kepler's laws because of the small masses in comparison to that of the Sun, and hence slightly improved upon Kepler's model by mathematical derivation. Kepler's third law of two-body problem can also be expressed in modern notations as \cite{2}:

\begin{equation}\label{eq1}
\displaystyle T|E|^{3/2}=\frac{\pi}{\sqrt{2}}Gm_1m_2\sqrt{\frac{m_1m_2}{m_1+m_2}}=\frac{\pi}{\sqrt{2}}G\bigg[\frac{(m_1m_2)^3}{m_1+m_2}\bigg]^{1/2},
\end{equation}
where $E$ is the total energy of the system that contains two gravitationally interacting point masses, $m_1$ and $m_2$, and $G$ is the gravitational constant. Kepler's third law has a wide range of applications, which range from the motion of artificial Earth satellites to the planets in the solar system, including the calculation of the stellar mass in the faraway universe.

There is a perfect analytical solution for the two-body problem. However, if there are more celestial bodies in the system, even one more? It would culminate in the three-body and n-body problems in celestial mechanics, which are age-old and challenging puzzles that Newton, Euler, Lagrange and Laplace have already studied. Until 1890, Poincar\'{e} determined that there is no analytical solution for a three-body problem in general, and its motion is usually non-periodic \cite{3}, which might explain why only five special solutions have been found under restricted conditions for over 300 years. These special solutions are now known as libration points or Lagrange points \cite{4}.

Benefiting from the substantial improvement of computer ability, Moore found the famous figure-eight periodic orbit for a three-body problem in 1993 \cite{5}. In recent years, \v{S}uvakov and Dmitra\v{s}inovi\'{c} made a breakthrough in finding 13 new distinct planar periodic orbits for the special three-body problem, containing three equal masses in a plane with zero angular momentum \cite{6}. Li and Liao successfully gained 695 families of periodic orbits in the same three-body system \cite{7} and furthermore, more than 1000 periodic orbits for a similar system with unequal masses were found by using a new numerical method in Ref. \cite{8}. Moreover, based on the statistical analysis of the derived periodic orbits, Dmitra\v{s}inovi\'{c} and \v{S}uvakov \cite{9}, Li and Liao \cite{7}, and Li et al. \cite{8} proposed that: similar to the elliptical motion of the two-body problem, there may be a relation in the form $T|E|^{3/2}=constant$ for the periodic motion of the three-body problem.

Does the three-body and/or n-body system have a similar Kepler's third law for the two-body system? Recently, Bohua Sun from the Cape Peninsula University of Technology in South Africa investigated this problem \cite{10} with the dimensional analysis, the method that Galileo and Newton had adopted, and for that which were widely recognized after Buckingham in 1914 \cite{11}. Bohua Sun chose the parameter of the gravitational field, $\alpha_n=Gm_im_j$ ($m_i$ and $m_j$ are the $i$th and $j$th point masses in an n-body system respectively), the reduced mass, $\mu_n$, and the area of the closed orbit, $A_n$, as the basic parameters in the dimensional analysis, and obtained the core relation of Kepler's third law of n-body periodic orbits: $T_n|E_n|^{3/2}=const.\times\alpha_n\sqrt{\mu_n}$, where $T_n$ is the orbital period, and $E_n$ is the total mechanical energy of the n-body system. The dimensions of these basic parameters $\alpha_n$, $\mu_n$ and $A_n$ are $L^3M\bar{T}^{-2}$, $M$ and $L^2$ respectively ($L$, $M$ and $\bar{T}$ refer to the basic dimensions of length, mass and time respectively). Then, substituting the core relationship into Eq. (1) and using Kepler's third law for the two-body problem, Bohua Sun determined that $const.=\frac{\pi}{\sqrt{2}}$, and proposed a complete conjecture on Kepler third law of the three-body periodic orbits by using the mass product symmetry of Newtonian gravitational field:
\begin{equation}\label{eq2}
\displaystyle T_3|E_3|^{3/2}=\frac{\pi}{\sqrt{2}}G\bigg[\frac{(m_1m_2)^3+(m_1m_3)^3+(m_2m_3)^3}{m_1+m_2+m_3}\bigg]^{1/2},
\end{equation}
and further extended to the n-body system:
\begin{equation}\label{eq2}
\displaystyle T_n|E_n|^{3/2}=\frac{\pi}{\sqrt{2}}G\bigg[\frac{\sum_{i=1}^n\sum_{j=i+1}^n(m_im_j)^3}{\sum_{k=1}^nm_k}\bigg]^{1/2},
\end{equation}
where $m_k$ is the $k$th point mass in the n-body system.

In order to verify the proposed conjecture, Bohua Sun compared the relation expression Eq. (2) with the ``period - energy'' fitting formulas, $T|E|^{3/2}=3.074m_3-0.617$, from the thousands of periodic orbits found by Li and Liao \cite{7} and Li et al. \cite{8}. As showed in Fig. 1, amazing consistency can be observed for $m_3>1$, while for $m_3<1$ there is a significant difference. But it is worth noting that Bohua Sun's relation expression Eq. (2) is consistent with the result of the two-body problem Eq. (1), when $m_3$ approaches to zero; while the result of Li et al. \cite{8} is negative and tends to be zero at $m_3=0.617/3.074$, those unreasonable results might come from the data fitting that should hence be revisited in future.

\begin{figure*}\footnotesize
\centering
\begin{tabular}{cc}
\includegraphics[scale=0.25]{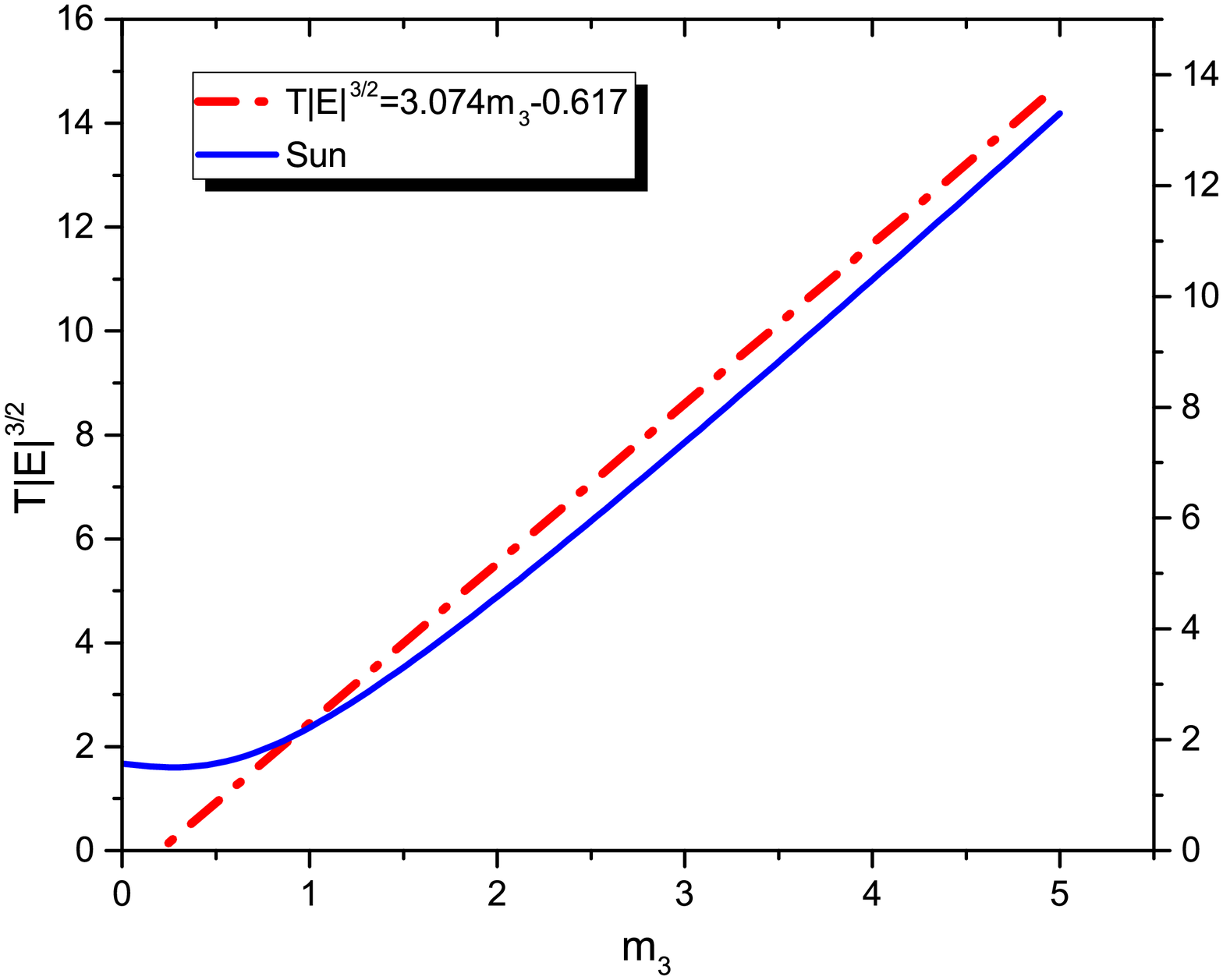} & \includegraphics[scale=0.25]{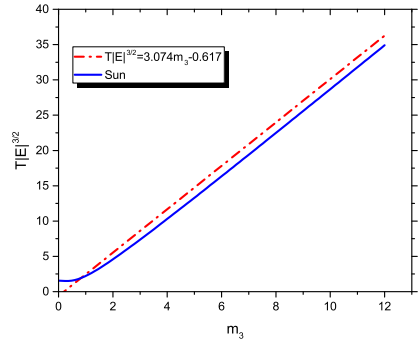} \\
(a) $m_3\in[0,5]$ & (b) $m_3\in[0,12]$ \\
\includegraphics[scale=0.25]{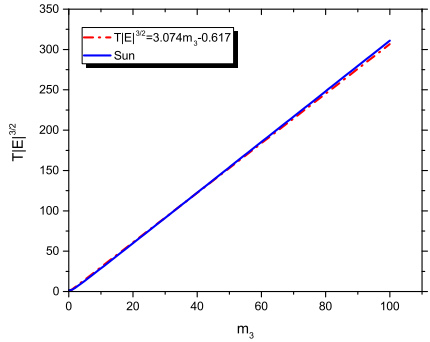} & \includegraphics[scale=0.25]{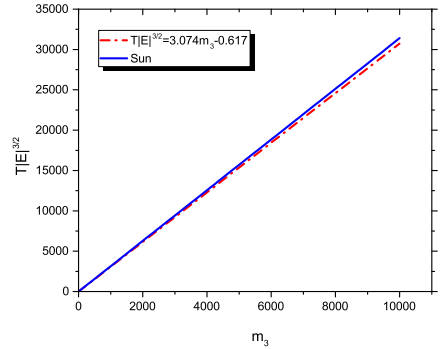} \\
(c) $m_3\in[0,100]$ & (d) $m_3\in[0,10000]$ \\
\end{tabular}
\caption{Comparing different $m_3$. Adapted from Ref. \cite{10}.}
\label{figure1}
\end{figure*}

From Kepler's third law of two-body elliptical motion Eq. (1), extended to the conjecture on Kepler's third law of three-body periodic motion Eq. (2), and then to the conjecture on Kepler's third law of n-body periodic motion Eq. (3), the relation expressions proposed by Bohua Sun are perfect and reasonable in mathematical form. Even if Bohua Sun's conjecture cannot be universally applicable to all periodic orbits, but only to a certain family of periodic orbits, it provides a shortcut in search of the periodic solutions of three-body and n-body problems and has valuable application prospects in space exploration. Recently, She also made some comments on Bohua Sun's conjecture \cite{12}.

Bohua Sun's conjecture on Kepler's third law of three-body and n-body problems \cite{10} is bound to arouse more attention towards the periodic solutions of three-body and n-body problems. Hence, further research should be conducted in this regard, which should include numerical simulation and strict theoretical deduction.

\section*{Acknowledgment}
\hskip\parindent We thank Professor Xin Wang for the valuable comments that helped to substantially improve the manuscript and thank Dr. Sheng-Xian Yu for the proofreading of the manuscript.


\begin{thebibliography}{0}
\bibitem{1} Plummer H C {1918 \emph{An Introductory Treatise on Dynamical Astronomy} (London: Cambridge University Press)}
\bibitem{2} Landau L D and Lifshitz E M {1976 \emph{Mechanics, 3rd ed.} (Oxford: Butterworth-Heinemann)}
\bibitem{3} Poincar\'{e} H {1890 \emph{Acta Math.} \textbf{13} 5}
\bibitem{4} Fitzpatrick R {2012 \emph{An Introduction to Celestial Mechanics} (Cambridge: Cambridge University Press)}
\bibitem{5} Moore C {1993 \emph{Phys. Rev. Lett.} \textbf{70} 3675}
\bibitem{6} \v{S}uvakov M and Dmitra\v{s}inovi\'{c} V {2013 \emph{Phys. Rev. Lett.} \textbf{110} 114301}
\bibitem{7} Li X M and Liao S J {2017 \emph{Sci. China-Phys. Mech. Astron.} \textbf{60} 129511}
\bibitem{8} Li X M, Jing Y P and Liao S J {2018 \emph{Publ. Astron. Soc. Japan} \textbf{70} 64}
\bibitem{9} Dmitra\v{s}inovi\'{c} V and \v{S}uvakov M {2015 \emph{Phys. Lett. A} \textbf{379} 1939}
\bibitem{10} Sun B H {2018 \emph{Sci. China-Phys. Mech. Astron.} \textbf{61} 054721}
\bibitem{11} Buckingham E {1914 \emph{Phys. Rev.} \textbf{4} 345}
\bibitem{12} She Z S {2018 \emph{Sci. China-Phys. Mech. Astron.} \textbf{61} 094531}

\end{thebibliography}
\end{document}